# A superconducting-nanowire 3-terminal electronic device

Adam N. McCaughan, Karl K. Berggren

*Electrical Engineering and Computer Science Department, Massachusetts Institute of Technology, 77 Massachusetts Avenue, Cambridge, MA 02139*

Superconducting classical and quantum devices have key advantages over conventional electronics, including reversible logic operations with almost no energy dissipation, and operating speeds exceeding 100 GHz[1]. In existing superconducting electronic systems, Josephson junctions[2] play a central role in processing and transmitting small-amplitude electrical signals. However, Josephson-junction-based devices have a number of limitations including: (1) sensitivity to magnetic fields, (2) limited gain, (3) inability to drive large impedances, and (4) difficulty in controlling the junction critical current (which depends sensitively on sub-Angstrom-scale thickness variation of the tunneling barrier). Here we present a nanowire-based superconducting electronic device, which we call the nanocryotron (nTron)[3], that does not rely on Josephson junctions and can be patterned from a single thin film of superconducting material with conventional electron-beam lithography. The nTron is a 3-terminal, "T"-shaped planar device with a gain of ~20 that is capable of driving impedances of more than 100 kΩ, and operates in typical ambient magnetic fields at temperatures of 4.2K. The device uses a localized, Joule-heated hotspot[4,5,6] formed in the gate to modulate current flow in a perpendicular superconducting channel. We have characterized the nTron, matched it to a theoretical framework, and applied it both as a digital logic element in a half-adder circuit, and as a digital amplifier for superconducting nanowire single-photon detectors pulses. The nTron has immediate applications in classical and quantum communications, photon sensing and astronomy, and its performance characteristics make it compatible with existing superconducting technologies. Furthermore, because the hotspot effect occurs in all known superconductors[7,8], we expect the design to be extensible to other materials, providing a path to digital logic, switching, and amplification in high-temperature superconductors.

The nanocryotron (nTron) is a two-dimensional superconducting device with a gate, a drain, and a source terminal, depicted in Fig. 1a. As shown in Fig. 1b, the gate terminal perpendicularly intersects the side of the channel via a narrow choke-point. Current entering the gate terminal, $I^{gate}$, switches the phase of the choke from the superconducting (S) to the resistive (R) state. The S→R phase transition is induced in the ~15-nm-wide choke by locally exceeding the critical current density $J_c$ of the niobium nitride film. In turn, the resistive phase of the choke induces a nonlinear suppression of the critical current of the channel, $I_c^{channel}$. The resulting dependence of the channel critical current on the gate input current, $I_c^{channel}(I^{gate})$, enables the nTron to produce robust switching and gain. Fig. 1c depicts the characterization of a non-inverting nTron amplifier circuit. The form of $I_c^{channel}(I^{gate})$ was ideal for a digital logic family: nearly zero modulation of the channel critical current was seen until a threshold of gate current, $I_c^{gate}$ = 2.9 µA, was reached. Exceeding that threshold produced a 30.5 ± 0.5% reduction in $I_c^{channel}$ from its base value $I_c^{channel}(0) = I_{c0}$. This reduction occurred coincidentally with a nonzero resistance measured at the gate terminal, indicating the formation of the resistive hotspot in the choke was responsible for the suppression of $I_c^{channel}$. Bias currents $\gtrsim 0.9\, I_{c0}$ resulted in undesired behavior such as photon- and noise-induced hotspot generation, while operating the devices at lower bias currents improved their robustness to source noise and typical ambient magnetic noise.

Despite the difference in underlying physical phenomena, the nTron is functionally analogous to a transistor. Fig. 1d depicts the digital operation of the nTron as it moves between three distinct states. Under the bias condition $I_c^{channel}(I_c^{gate}) < I^{bias} < I_{c0}$, when a logical LOW ($I^{gate} < I_c^{gate}$) is fed into the gate input, the channel remains superconducting, and when a logical HIGH ($I^{gate} > I_c^{gate}$) is input, the channel becomes resistive. In this context, we used the nTron as a discrete digital element for performing logical functions. Digital operation was reproduced for load impedances of 50 Ω, 100 Ω, 1 kΩ, 10 kΩ, 100 kΩ, and the open-circuit case. In the open-circuit case, the bias condition of $I^{channel} = 0.85\ I_{c0} = 90$ μA yielded an output voltage of 8.1 V and an input-output isolation of 42.7 kΩ.

Fundamental to the operation of the nTron is the phenomenon of localized critical-current suppression, wherein a hotspot sustained by Joule heating suppresses the superconducting characteristics of the nearby material. Phonons and quasiparticles diffuse from the hotspot to the surrounding superconductor where they interact with the superconducting bath, depleting the local Cooper pair population as they relax back to equilibrium[9]. In the case of thin-film NbN, out-diffusion of hot electrons is the primary means of thermal energy transfer from the hotspot to the surrounding material[10,11]; the characteristic diffusion coefficient for non-equilibrium electrons has been measured[12] to be 45 nm$^2$/ps.

To corroborate our measurement results and facilitate future designs, we simulated the nTron device geometry using an established theoretical framework, the two-temperature model[13], which uses an effective electron temperature $T_e$ to represent the temperature of populations of quasiparticles and Cooper pairs. The simulation used no free parameters—we instead employed measurements from the device as well as empirical parameters for thin-film NbN found in the literature[11,12]. The simulation results showed that when a hotspot was formed in the choke of the

nTron, the effective electron temperature $T_e$ of the surrounding ~100 nm was increased. This temperature increase corresponds to a decrease in $J_c$ over the same radius, effectively reducing the total channel critical current. As a result of the hotspot formation, the $I_{c0}$ was reduced by 28% of its original value, closely matching the measured value of 30.5%. Additionally, just above $I_c^{gate}$ the resistance of the simulated gate hotspot was 823 Ω, which was comparable to the measured resistance of 832 Ω.

The nTron consists of a single layer of thin-film superconducting material, so its 2D geometry defines its operation. The layout of the device has several essential design elements, resulting in a large design parameter space in which future implementations may be optimized. The most critical design element is the size and location of the choke region, which is the point of highest current density for the gate input current and is where the hotspot first forms $I^{gate} > I_c^{gate}$. The choke hotspot is the key to the nTron's functionality; it proximitizes the nearby superconducting channel and induces a suppression of the critical current in that area. The width of the choke defines the input current level required for the gate hotspot to form and produce a logical HIGH, which can be approximated by $I_c^{gate} = J_c d\, w_{choke}$, where $J_c$ is the critical current density, $d$ is the substrate thickness, and $w_{choke}$ is the width of the choke. As we observed in our simulation, the formation of the hotspot only suppressed $J_c$ within approximately one diffusion length $L_D = \sqrt{D_e \tau_r}$ (~100 nm for thin-film NbN) of its perimeter. Accordingly, the width of the channel must be on the same length scale for the hotspot to generate a sharp dropoff in $I_c^{channel.}$.

To demonstrate the suitability of the nTron for digital applications, we used it to build AND/OR/NOT/COPY gates (Fig. 2a), and from those constructed a half-adder (Fig. 2b). We operated the half-adder in a pipelined fashion, where valid inputs were translated into valid outputs for the next stage only upon the enabling of the gate bias current. After the computation was completed and the final outputs were recorded, all the input and bias currents were shut off to unlatch the gates and reset the computation. For the purposes of this demonstration, operating in the latching regime enabled us to tolerate potentially large fabrication defects in these first devices. In the latching regime, variations between the OFF-state input impedances of the gates did not matter because each stage was able to drive arbitrarily large input impedances in the next stages—to operate in a non-latching regime, output impedances must be more carefully controlled[14]. No electrical or magnetic shielding was necessary.

We performed a number of additional characterization experiments to demonstrate RF operation, sensitivity, and robustness of the nTron. To characterize the device at higher frequencies, we input a 10 MHz square wave into the device to generate an eye diagram. With a 1.46 kΩ output load, the nTron was able to convert a 3.10 ± 0.02 µA input square wave into a 62.7 ± 1.2 µA output square wave, corresponding to a signal gain of 20.2. At the sampling point, the signal-to-noise ratio was 168. Applying a magnetic field perpendicular to the device plane, swept between ±7.4 mT, had no measureable impact to the eye diagram characteristics. Details of these experiments are available in the Supplementary Information. Additionally, we measured the input level sensitivity by operating the device as a current comparator, measuring an $I_c^{gate}$ of 2.91 µA with a 1-σ grey zone of 66 nA.

We tested the nTron's ability to amplify pulses as well as established an upper bound to its jitter by integrating it with a superconducting nanowire single-photon detector[15] (SNSPD). When detecting a photon, SNSPDs produce millivolt-scale microwave pulses with sub-100 ps rising edges. The integrated device used these pulses as input into the gate of a nTron amplifier. We monolithically fabricated the SNSPD and nTron on the same film within a 100 µm² area, connecting the SNSPD output to the gate of the nTron. Our circuit design (Supplementary

Fig. S2) enabled us to bias each device separately, as well as simultaneously read out the unamplified SNSPD pulses and nTron-amplified pulses for comparison. When the SNSPD was illuminated with a 1550 nm sub-picosecond laser, output pulses were produced from both devices concurrently. We compared the SNSPD output to the nTron-amplified output, and observed a factor of 2.9 increase in signal pulse amplitude. This increase in amplitude proportionally increased the slew rate of the rising edge, resulting in a reduced jitter when measured by our scope. Jitter was measured between the sync edge of the sub-ps laser and the rising edge of each device's output pulses. The unamplified pulses had a full-width half-max jitter of 41.3 ± 0.3 ps, while the corresponding nTron-amplified pulses showed a reduced full-width half-max jitter of 23.8 ± 0.2 ps. This value of 23.8 ps also sets an upper bound to the nTron input-to-output jitter, although the actual value should be significantly smaller, as state-of-the-art NbN SNSPDs[16] alone produce jitters on the order of 20-40 ps. Details of the SNSPD experiment and measurement results are available in the Supplementary Information.

The central result of this report is the design and characterization of a superconducting three-terminal device that does not rely on Josephson junctions. The device was used to demonstrate digital operations, as well as amplification with picosecond timing resolution. The ease of fabrication and ability to drive large-impedance loads suggest a broad utility in applications ranging from read-out electronics for superconducting sensors to high-performance computing.

**Methods Summary**

All the nTron devices used in this paper were fabricated from a contiguous ~10 nm film of niobium nitride (NbN) deposited on a single 2" R-plane sapphire wafer. For the simulation, parameters extracted from the device and the film from which it was fabricated included (1) the device critical current density $J_c = 6.6$ MA/cm$^2$, (2) the critical temperature $T_c = 12.6$ K and its transition width $\Delta T_c = 1.1$ K of the film, and (3) the device sheet resistance $R_s = 285$ Ω/sq. For patterning, we spun on ~50 nm hydrogen silsesquioxane (HSQ) resist and exposed the device patterns in a 125 kV Elionix electron-beam lithography tool. We then etched the NbN around the patterned HSQ to complete the nTron fabrication. Contact pads were added by photolithography of 1-μm-thick Shipley S1813 photoresist, followed by evaporation of 10 nm Ti and 25 nm Au and then liftoff. Electrical connections between the sample mount and device contact pads were made using aluminum wirebonds. For all experiments described here, samples were submerged in a bath of liquid helium. Experiments took place in ambient magnetic conditions. Further detail on the individual experiments may be found in the Supplementary Information.

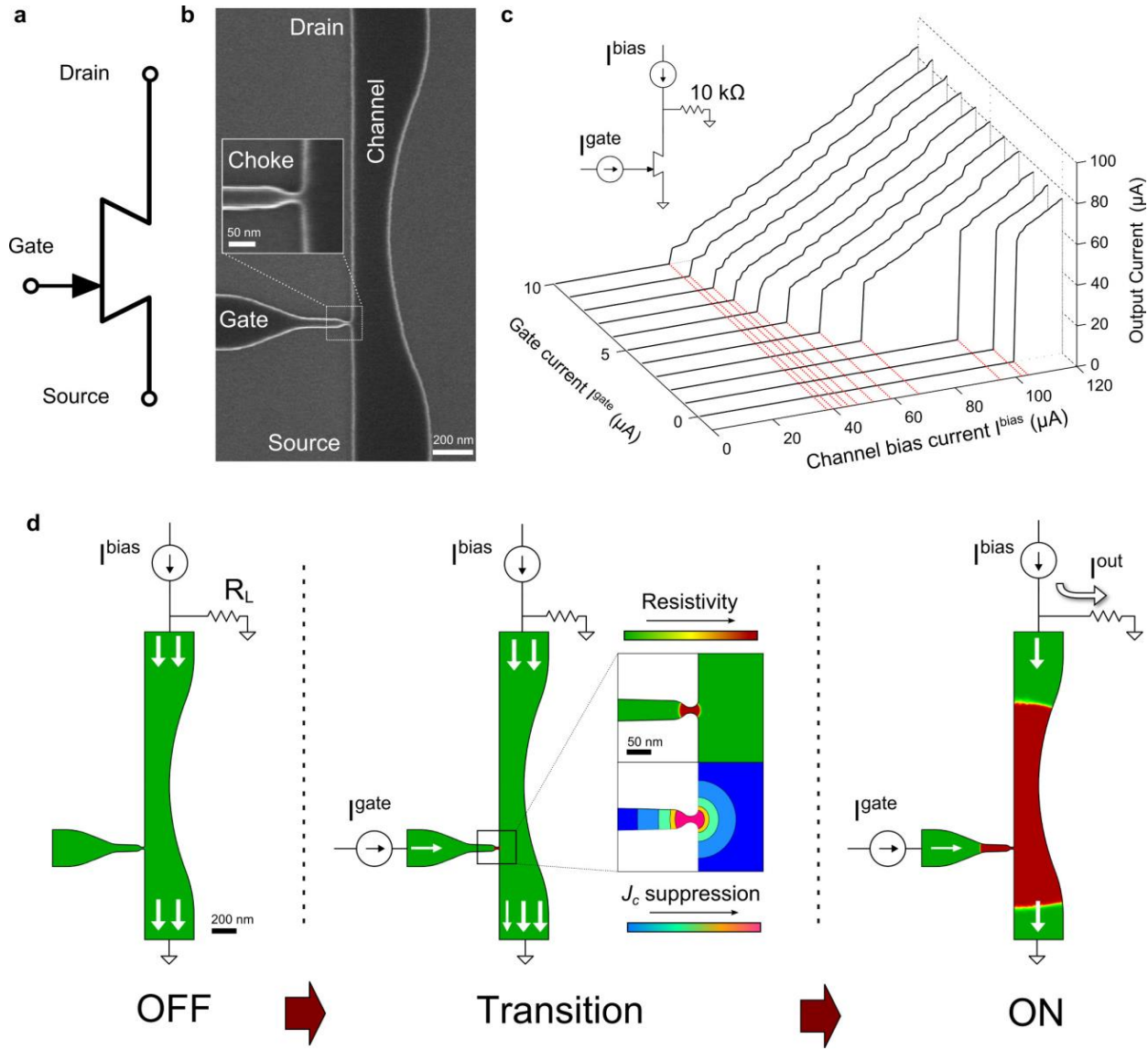

**Figure 1 | Basic operation of the nanocryotron (nTron). a**, Three-terminal circuit symbol. The position of the gate arrow denotes the location of the choke relative to the narrowing of the channel. **b**, SEM of a fabricated nTron, the inset depicts a close-up of the choke, the area in which the resistive hotspot is first formed. **c**, Circuit schematic and output characteristics for an nTron in a noninverting amplifier configuration. $I^{gate}$ was fixed and $I^{bias}$ was swept from 0 to 120 μA. **d**, Numerical simulation of the nTron depicting the three states of operation. OFF: The device is fully superconducting, bias current is drained through the channel to ground. Transition: Current is added to the gate input, forming a resistive hotspot which locally suppresses superconductivity. (inset, upper) Closeup of the resistive hotspot forming in the

choke. (inset, lower) Contour map of $J_c$ suppression extending from the hotspot. From inner to outer, the bands represent reductions in $J_c$ by 99% (magenta), 75% (orange), 50% (green), 25% (light blue), and 0% (blue). ON: The critical current of the channel is reduced sufficiently that the bias current triggers the formation of a resistive hotspot in the channel.

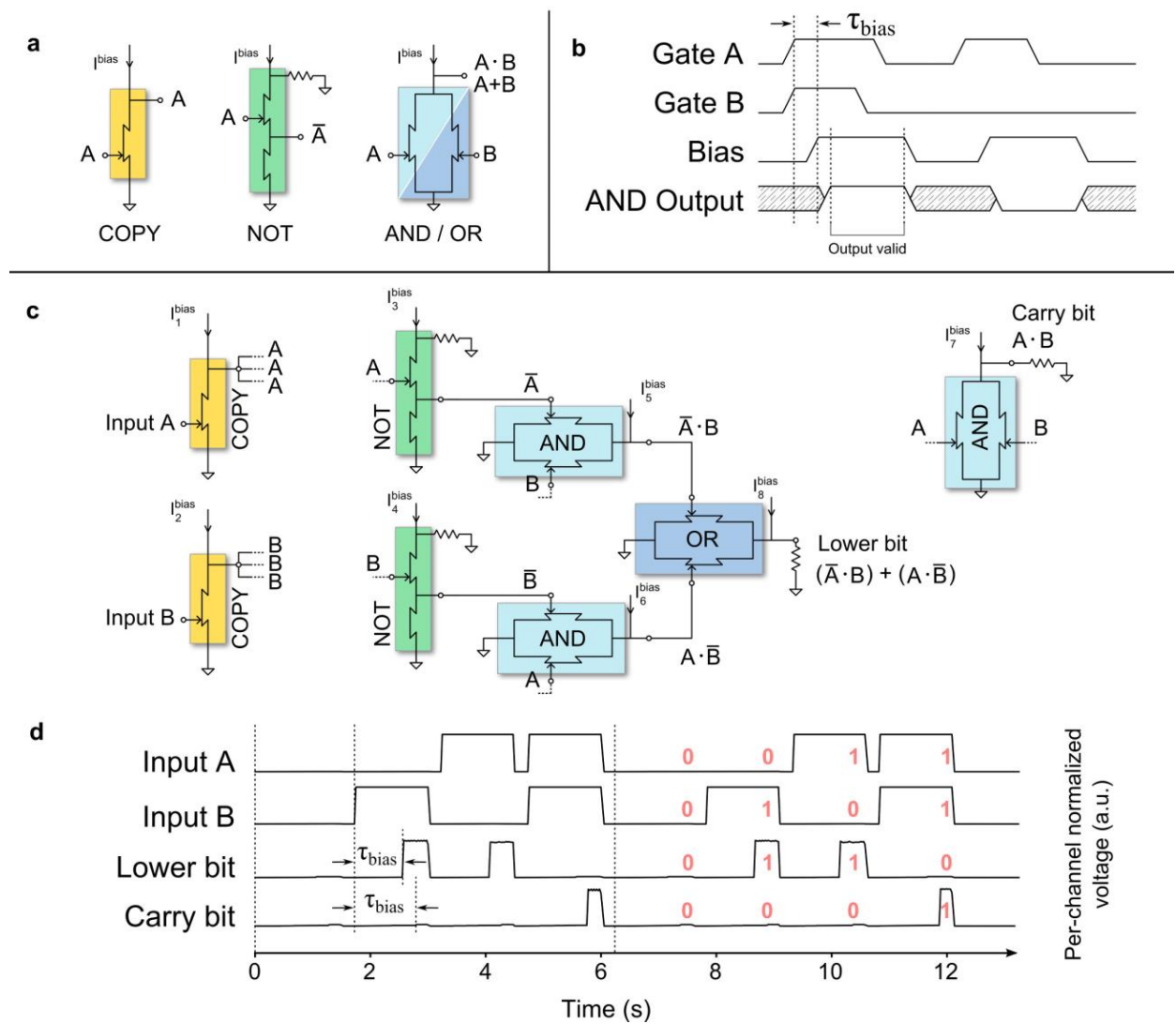

**Figure 2 | Digital gates based on the nanocryotron, and demonstration of a half-adder. a**, Schematic of a set of universal logical gates from the basic three-terminal nTron. The AND gate and OR gate are topologically identical, and are only differentiated by their bias conditions. AND/OR/COPY were constructed purely from nTrons, while the NOT gate required a shunt impedance for the bias (in this case a resistor). See Supplementary Information for a description of each gate's operation. **b**, AND-gate timing diagram for pipelined logic propagation. Once

gates A and B have valid inputs, the bias current is enabled and the resulting output can be used as an input for the next stage. $\tau_{bias}$ denotes the propagation delay due to the low-rate bias electronics. **c**, Half-adder circuit schematic constructed from logical gates. Single inputs were provided into the initial (yellow) COPY gates, which acted as a buffer to fan the *Input A* and *Input B* signals out to the other gates. **d**, Per-channel output for the half-adder for computation of 0+0, 0+1, 1+0, and 1+1, repeated twice. HIGH (1) and LOW (0) current values were input to *Input A* and *Input B*, and after a bias electronics delay $\tau_{bias}$, the lower bit and carry (upper) bit outputs represented the resulting sum of the inputs. The red text overlay of ones and zeros corresponds to HIGH and LOW values.

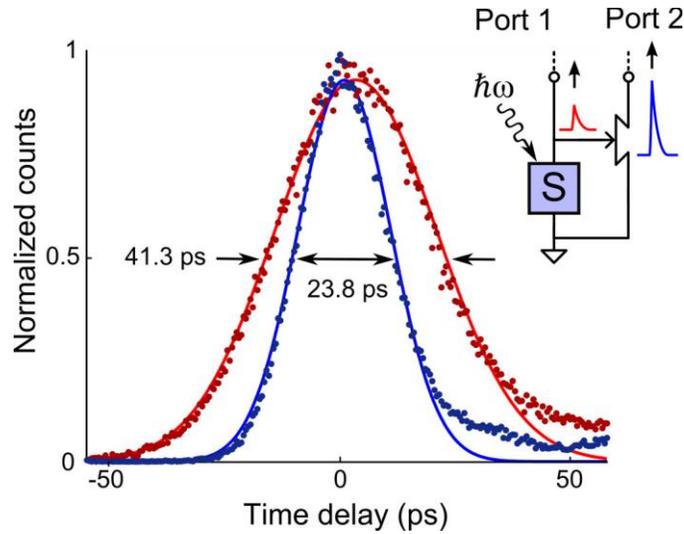

**Figure 3 | Jitter measurements for an nTron integrated as an amplifier for a superconducting nanowire single-photon detector (SNSPD) pulses.** Detection of laser photons from a sub-ps laser by the detector (inset, purple 'S' box) generated an electrical pulse on Port 1 (inset, red) and also triggered a concurrent, amplified pulse from the nTron on Port 2 (inset, blue). Plotted is a histogram of the relative delay between the laser sync edge and the resulting electrical pulse edges of the unamplified SNSPD (red dots) and nTron-amplified output (blue dots). Gaussian fits to each data set are shown as solid lines. The reduced jitter in the amplified signal is due to increased signal amplitude. (upper right) Device schematic of the integrated SNSPD-nTron pulse amplifier. Full circuit schematic and description available in Supplementary Information.

**Acknowledgements** The authors would like to thank Faraz Najafi, Francesco Bellei, Andrew Dane, and Yachin Ivry for helpful discussions. Additionally, they would like to thank James Daley and Mark Mondol for nanofabrication technical support. This work was supported by the National Science Foundation (NSF). Adam McCaughan was supported by a fellowship from the NSF iQuISE program, award number 0801525.

# Supplementary Information

**History:**

Our device is based on and inspired by attempts in the late 1950s to develop superconducting logic systems (prior to the invention of the Josephson junction) that used electromagnetic fields to suppress superconductivity in a nanowire[1]. These four-terminal devices were termed "cryotrons" and were abandoned after the development of the Josephson junction and the SQUID. In the intervening years, Josephson-junction-based technologies have dominated the literature, but a number of other devices have also been introduced. These include tunable-supercurrent SNS junctions[2], resistive heaters stacked on top of superconducting films[3], and Josephson field-effect transistors[4]. Despite their diversity, all of these devices have required two or more active layers, and none have been demonstrated beyond the characterization of their basic three or four-terminal unit. Recently, it was proposed[5] that building a 1950s-type cryotron at the nanoscale would offer similar power and clock performance to that of state-of-the-art Josephson-junction-based devices. However, fabricating such a device would likely require complex fabrication procedures: integration of multiple superconducting materials, extremely thin insulating barriers, and two or more active layers. The closest antecedent to the device presented here was an amplifier for superconducting single-photon detectors developed by Ejrnaes et al.[6] It used a simpler, but less robust, amplification mechanism, and it too was never developed beyond its use as a pulse amplifier. In homage to the cryotron origins of the cryotron concept, we thus named our device the nanocryotron, or nTron.

**Extended nTron logic gate description:**

Fig. 2a shows the COPY/NOT/AND/OR gates. All of the gates required only one or two nTrons, except for the NOT gate which used an additional 330 Ω shunt resistor and another nanowire constriction in addition to a single nTron. The COPY gates operate similar to a non-inverting version of the FET: when a logical HIGH ($I^{gate} > I_c^{gate}$) is input to the gate, the channel becomes resistive and the channel bias current is diverted into the output, generating a HIGH output. When a logical LOW ($I^{gate} < I_c^{gate}$) is input to the gate, the channel bias current drains directly to ground without generating a resistive region in the channel. As a result, the output is effectively shorted to ground and is thus a logical LOW. The AND and OR gates are the result of putting two nTrons in parallel. In the case of the OR configuration, when either gate input received a logical HIGH, the combined $I_c^{channel}$ of the two nTrons would be reduced enough such that the channel bias current would switch the device. The AND configuration was achieved by biasing the channel below this point, to the point where the combined channel critical currents only dropped below the bias current when both gate inputs were HIGH. This topological equivalence between the AND and OR gate opens the possibility to dynamically reprogram an otherwise fixed logic circuit.

For the NOT gate, when a logical LOW was input, $I_c^{channel}$ was greater than the bias current, allowing all the current to pass through it without switching. However, $I_c^{pulldown}$ was designed to be smaller than the full channel bias current, which caused the pulldown constriction to switch and diverted the excess current into the output, producing a logical HIGH out. In the case of a HIGH input, the suppressed $I_c^{channel}(I^{gate})$ was less than the bias current. As a result, the nTron transitioned to the ON state, and current was diverted to ground through the shunt resistor. The pulldown constriction served to tie the output to ground, even in the presence of the unavoidable small current leaking through the resistive channel.

**Experimental details: Half-adder**

The half-adder experiment consisted of fabricating several individual gates on a single chip, connecting them via wirebond, and biasing them in a pipelined manner to perform the summing computation. We built a custom 16-channel combined ADC and DAC system to handle the multiple gate and bias inputs, and to read out their bias voltages. For the gate inputs and channel biases, current sources were approximated using the DAC voltage channels in series with 100 k$\Omega$ resistors. The ADC channels were used to record the status of each gate's output. Due to the pipeline-propagation of the half-adder, in combination with the low-rate DAC/ADC system, the computation of the half-adder outputs required approximately 0.8 seconds to complete. No external amplification was necessary for readout, and the lower bit and carry bit and generated 38.8 mV and 17.0 mV respectively across their 330 $\Omega$ load resistors. The sampling rate of the room-temperature electronics setup limited the computation rate to 1.2 Hz. In this mode, longer time scales present more opportunity for noise to erroneously switch the devices. Despite this disadvantage, exercising the circuit over 4,000 cycles spanning 55 minutes the circuit produced only seven errors.

**Experimental details: 10 MHz eye diagram circuit**

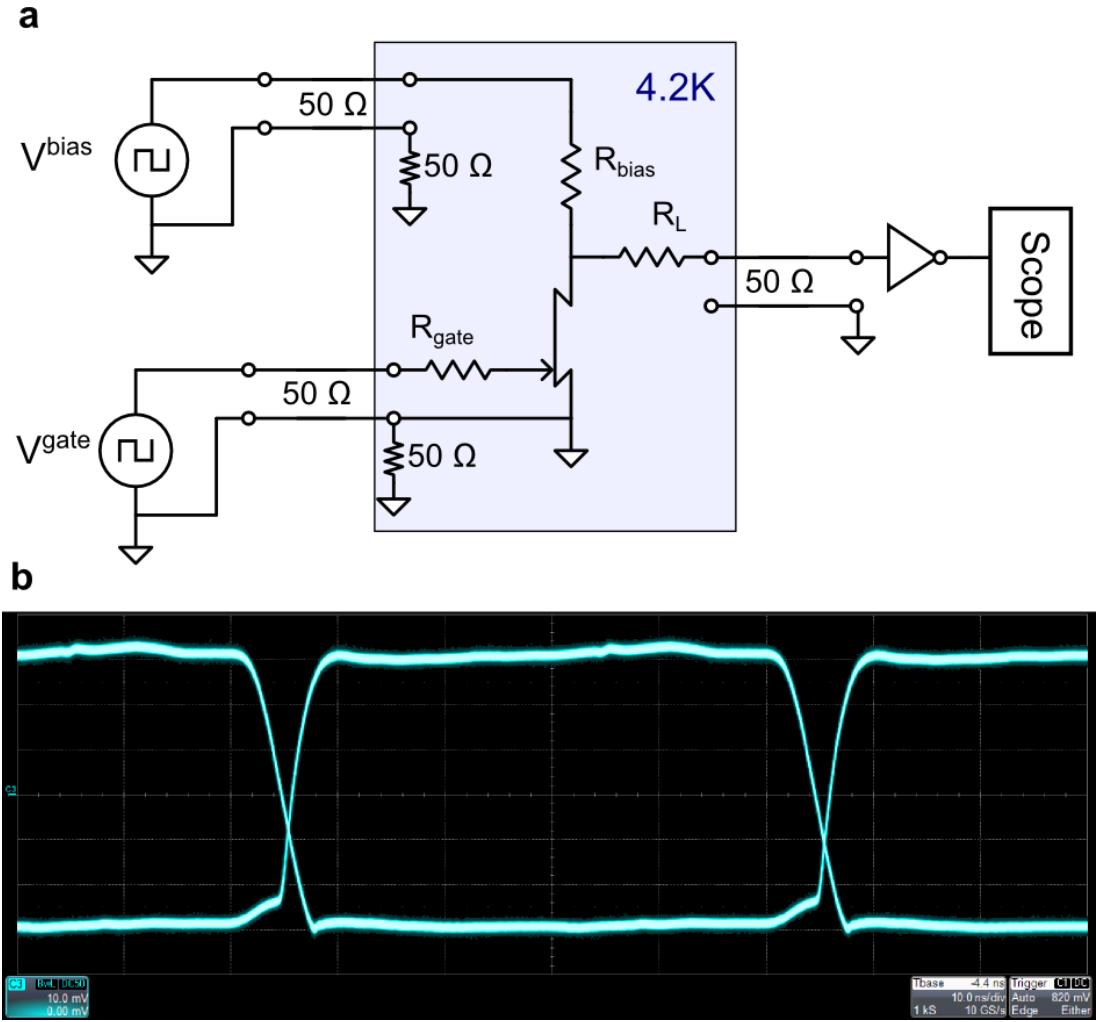

**Figure S1 | Circuit schematic for 10 MHz eye diagram experiment. a**, Circuit diagram for the nTron 10 MHz eye-diagram experiment. The area in blue represents the portion of on the sample holder and submerged in liquid helium at 4.2 K. Placing the resistors close to the device allowed us to convert the incoming voltage square waves to a low-amplitude current square waves. The resistors $R_L$, $R_{bias}$, and $R_{gate}$ were 1.46 kΩ, 20.8 kΩ, and 42.0 kΩ, respectively (as measured at 4.2 K).

Using the experimental setup shown in Fig. S1a, we generated a 10 MHz eye diagram for the

nTron digital amplifier. We accomplished this by sending one current square wave into the device channel, and one current square wave into the gate port (delayed by approximately 10 ns relative to the channel square wave). The channel square wave primed the nTron to switch from the ON to the OFF state, such that when the rising edge of the gate square wave arrived, the nTron switched, generating an output current which was read out by the scope.

**Experimental details: nTron comparator**

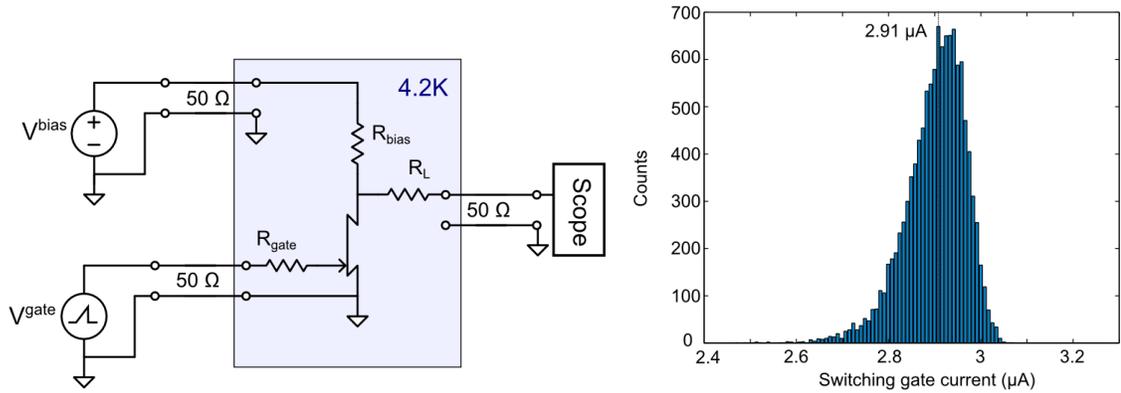

**Figure S2 | The current comparator experiment. a**, Circuit diagram for the nTron current comparator. The channel was biased at a fixed value, and the gate was ramped until output appeared at the scope. **b,** Histogram of $I^{gate}$ values for the gate current at which the comparator switched and produced an output voltage at the scope.

We tested the nTron as a comparator to characterize the current sensitivity of the gate input. As shown in Fig. S2, the experimental setup was the same as that of the 10 MHz eye diagram characterization, only differing in the current bias and readout scheme. First, we applied a 52.8 µA bias current to the channel in order to prime it for transition from the ON to OFF state when $I^{gate}$ exceeded $I_c^{gate}$. We then ramped the gate current from 0 to 15 µA at a rate of 16.7 nA/µs, recording the output current through the load resistor $R_L$ on a 1 GHz oscilloscope. We repeated this experiment 11,000 times, plotting the results as a histogram (Fig. S2b). From the histogram, we found a mode switching value of 2.91 µA with a 1-σ grey zone of 66 nA.

**SNSPD + nTron amplifier circuit design/schematic and experimental details**

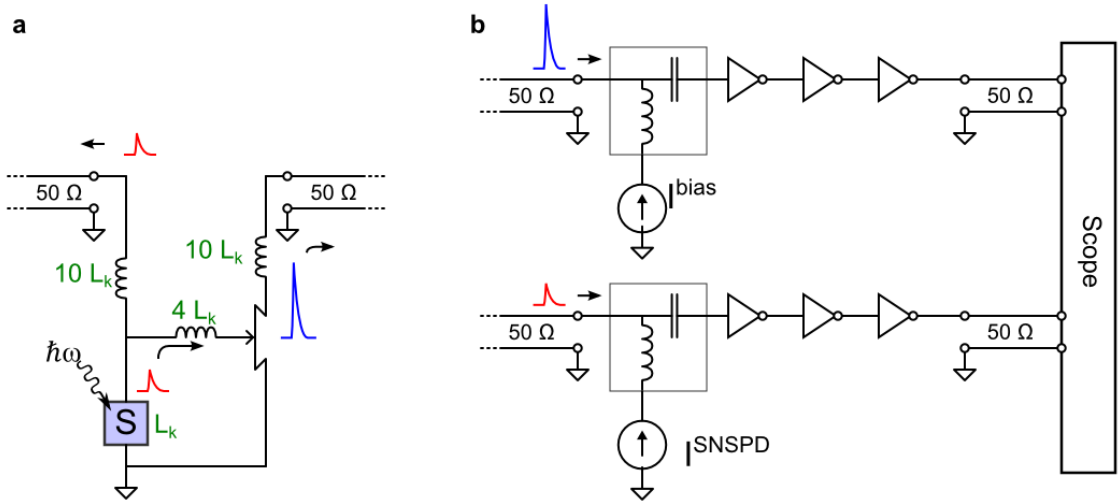

**Figure S3 | Circuit schematic for the SNSPD and nTron pulse amplifier experiment.** **a**, Device circuit schematic. The inductors were made by patterning long nanowires, which intrinsically produce kinetic inductance. The length of the inductor nanowires (and thus their total inductance) were scaled against the SNSPD, which had an approximate kinetic inductance of $L_k \approx 25$ nH. **b**, Room-temperature readout and bias electronics. Pulses generated from the device and output to the coax in (a) arrived at the other end of the coax, shown in (b), where they were amplified with 3× 20-3000 MHz amplifiers before being input to the scope.

As shown in the Fig. S3, the SNSPD and nTron pulse amplifier were integrated as single circuit. Current biasing was accomplished through the use of inductive splitting, where the inductance was provided by the kinetic inductance of the nanowires. The SNSPD nanowires had a width of 60 nm, and the inductor nanowires had widths of 200 nm. The entire circuit occupied an area of approximately 100 μm². The 50 Ω lines were high-frequency coaxial cable running between the sample and room-temperature electronics. During operation, we biased the SNSPD at 35 μA and the nTron at 95 µA. Expected operation was ensured by measuring: photon sensitivity for the SNSPD and amplifier when biased separately; critical current suppression in the nTron channel when the SNSPD was overbiased (creating a hotspot in the gate); count rate from both outputs when biased together; count rate vs. $I^{bias}$; and count rate from the SNSPD vs $I^{SNSPD}$. The results of these measurements allowed us to conclude that (1) the nTron amplifier, at a bias of 95 µA, was not photosensitive; (2) there was a 1:1 correspondence in counts between the two outputs (one amplifier pulse per SNSPD pulse); (3) no counts were generated in the amplifier when only the SNPSD was biased, and vice-versa.